\documentclass[aps,prd,twocolumn,tightenlines,nofootinbib,superscriptaddress]{revtex4-2}
\usepackage[utf8]{inputenc}
\usepackage{graphicx}
\usepackage{subcaption}
\usepackage{mwe}
\usepackage{mathrsfs}
\usepackage{amsfonts}
\usepackage{setspace}
\usepackage{cellspace}
\usepackage{amsmath,amssymb,bm}
\usepackage[colorlinks=true,linkcolor=blue]{hyperref}
\usepackage{xcolor}
\usepackage{epsfig}
\usepackage{slashed}
\usepackage{caption}
\usepackage{hhline,multirow,tabularx}  
\usepackage{dcolumn}    
\usepackage{url}        
\usepackage{braket}     
\renewcommand{\bra}[1]{\left<#1\left|}
\renewcommand{\ket}[1]{\right|#1\right>}

\begin{document}

\title{\texorpdfstring{Updated analysis of near-threshold heavy quarkonium production \\
for probe of proton's gluonic gravitational form factors}{Updated analysis of near-threshold heavy quarkonium production for probe of proton's gluonic gravitational form factors}}

\author{Yuxun Guo}
\email{yuxunguo@umd.edu}
\affiliation{Department of Physics, University of Maryland, College Park, MD 20742, USA}
\affiliation{Nuclear Science Division, Lawrence Berkeley National Laboratory, Berkeley, CA 94720, USA}

\author{Xiangdong Ji}
\email{xji@umd.edu}
\affiliation{Department of Physics, University of Maryland, College Park, MD 20742, USA}

\author{Yizhuang Liu}
\email{yizhuang.liu@uj.edu.pl}
\affiliation{Institute of Theoretical Physics,
Jagiellonian University, 30-348 Kraków, Poland}

\author{Jinghong Yang}
\email{yangjh@umd.edu}
\affiliation{Department of Physics, University of Maryland, College Park, MD 20742, USA}

\begin{abstract}
There has been growing interest in the near-threshold production of heavy quarkonium which can access the gluonic structure in the nucleon. Previously~\cite{Guo:2021ibg} we studied this process with quantum chromodynamics (QCD) and showed that it can be factorized with the gluon generalized parton distributions (GPDs) in the heavy quark limit. We further argued that the hadronic matrix element is dominated by its leading moments corresponding to the gluonic gravitational form factors (GFFs) in this limit. Since then, there have been many new developments on this subject. More experimental measurements have been made and published, and the lattice simulation of gluonic GFFs has been improved as well. In this work, we make an important revision to a previous result and perform an updated analysis with the new inputs. We also study the importance of the large momentum transfer to extract these gluonic structures reliably in this framework.

\end{abstract}
\maketitle

\section{Introduction}

The gluonic structures have been an important topic in quantum chromodynamics (QCD) and hadronic physics. On the one hand, gluons that mediate the strong interaction play a prominent role at the non-perturbative scale in the hadron. On the other hand, they are free of electroweak interaction and  much harder to probe than quarks. Consequently, there has been rising interest in the exclusive electro/photo-productions of heavy quarkonium. Assuming suppressed contributions from the intrinsic heavy quarks, these processes are dominated by the exchange of gluons and thus can be used to access the gluonic structures. Experiments with such purposes are planned in the future Electron-Ion Collider (EIC)~\cite{Accardi:2012qut}, whereas at Jefferson Laboratory (JLab) the near-threshold production of $J/\psi$ that requires less energy has been proposed and measured which revealed exciting results~\cite{Joosten:2018gyo,GlueX:2019mkq,Duran:2022xag,Adhikari:2023fcr}.

There have been many theoretical developments in the literature to analyze this process~\cite{Voloshin:1978hc,Gottfried:1977gp,Appelquist:1978rt,Bhanot:1979vb,Kharzeev:1995ij,Kharzeev:1998bz,Kharzeev:2021qkd,Gryniuk:2016mpk,Gryniuk:2020mlh,Hatta:2018ina,Hatta:2019lxo,Mamo:2019mka,Mamo:2021krl,Mamo:2022eui,Sun:2021pyw,Sun:2021gmi}. In the previous work of three of the authors~\cite{Guo:2021ibg}, we showed that the near-threshold photoproduction of heavy quarkonium can be factorized with gluon generalized parton distributions (GPDs), extending the factorization proved for the  diffractive production in the collinear limit~\cite{Collins:1996fb,Ivanov:2004vd}. Utilizing the heavy quark limit, we also argued that the hadronic matrix element will be dominated by the leading moments that correspond to the gluonic gravitational form factors (GFFs) near the threshold. These GFFs carry important information about the nucleon such as their mass, angular momentum and mechanical properties~\cite{Ji:1994av, Ji:1996ek, Polyakov:2002yz, Polyakov:2018zvc,Ji:2021mtz,Ji:2022exr,Burkert:2023wzr}. Although a complete determination of the gluonic GFFs from these measurements alone is still model-dependent~\cite{Duran:2022xag}, it provides us with an effective tool to handle such problems.

In this work, we perform an updated analysis of the near-threshold $J/\psi$ production following the previous work for two main reasons. First, we note that there was a missing factor of 2 in the hadronic matrix element $G(\xi,t)$ in ref.~\cite{Guo:2021ibg} due to a mix of conventions. Besides, there have been many new developments on this subject ever since our previous work. More data have been measured and published  recently by the $J/\psi$ 007 experiment at JLab Hall C~\cite{Duran:2022xag} and the GlueX collaboration at Hall D~\cite{Adhikari:2023fcr}. Moreover, the lattice QCD simulation of gluonic GFFs has been improved~\cite{Pefkou:2021fni}. The large $\xi$ expansion essential for relating this process to the gluonic GFFs requires careful treatments. Consequently, we will present more detailed analyses of these new results in this work.

\section{Revised formula}

We start by noting that the definition of gluon GPDs in eq.~(16) of the previous work~\cite{Guo:2021ibg} includes an extra factor of $\frac{1}{2}$ from the trace operator $\text{Tr}$, which should be removed. Correspondingly, the extra factor of $\frac{1}{2}$ in eq. (24) in ref.~\cite{Guo:2021ibg} shall be removed as well. The revision will leave the main cross-section formula unchanged, i.e., eqs.~(13) and (17) in ref.~\cite{Guo:2021ibg}. However, one needs to substitute the gluon GPDs with the corrected definition that reads,
\begin{align}
\label{gluonGPD}
   &F_g(x,\xi,t) \equiv\nonumber \\ &\frac{1}{(\bar P^+)^2} \int \frac{\text{d}\lambda}{2\pi} e^{i\lambda x}\bra{P'}F^{a+i}_{\;\;\;\;\;}\left(-\frac{\lambda n}{2}\right)F^{a+}_{\;\;i}\left(\frac{\lambda  n}{2}\right) \ket{P}\ ,
\end{align}
where the index $a$ sums over all colors. The $F_g(x,\xi,t)$ can be parameterized as~\cite{Ji:1996ek,Diehl:2003ny}
\begin{align}
    F_{g}(x,\xi,t)=\frac{1}{2\bar P^+}\bar u(P')\left[H_g \gamma^+ +E_g\frac{i\sigma^{+\alpha}\Delta_{\alpha}}{2 M_{N}}\right]u(P) \ ,
\end{align}
where $H_g$ and $E_g$ are the well-known $H_g(x,\xi,t)$ and $E_g(x,\xi,t)$ GPDs.
The hadronic matrix element $G(\xi,t)$ remains to be~\cite{Guo:2021ibg}
\begin{align}\label{eq:amplitude2}
G(t,\xi)=\frac{1}{2\xi} \int_{-1}^1\text{d}x{\cal A}(x,\xi)F_g(x,\xi,t) \ ,
\end{align}
where the Wilson coefficient ${\cal A}(x,\xi)$ reads
\begin{equation}
   {\cal A}(x,\xi)\equiv\frac{1}{x+\xi-i0}- \frac{1}{x-\xi+i0}\ .
\end{equation}

On the other hand, the extra factor of $\frac{1}{2}$ does affect the relations between the gluon GPDs and GFFs. Removing the extra factor of $\frac{1}{2}$ in the gluon GPDs causes an extra factor of 2 to be multiplied to each gluonic GFF when expanding the GPDs in terms of their moments. Consequently, in the unpolarized case we have
\begin{align}\label{gt2ff}
&|G(t,\xi)|^2=\frac{4}{\xi^4}\Bigg{\{}\left(1-\frac{t}{4M_N^2}\right)E_2^2 \nonumber \\ &-2 E_2(H_2+E_2) +\left(1-\xi^2\right)(H_2+E_2)^2 \Bigg{\}}\ ,
\end{align}
where an extra factor of $4$ appears accordingly, compared to the eq. (26) in ref.~\cite{Guo:2021ibg}. The same factor of $4$ should be multiplied to the polarized $|G(t,\xi)|^2$ in eqs. (58) and (59) in ref.~\cite{Guo:2021ibg} as well. Recall that the GFFs $H_2\equiv H_2(t,\xi)$ and $E_2\equiv E_2(t,\xi)$ follow the same definition as ref. \cite{Guo:2021ibg} that reads
\begin{align}\label{eq:E0}
    \int_{0}^1 dx H_g(x,\xi,t)=A_{2,0}^g(t)+(2\xi)^2C^g_{2} \equiv H_2(t,\xi)  \ , \nonumber \\
    \int_{0}^1 dx E_g(x,\xi,t)=B^g_{2,0}(t)-(2\xi)^2C^g_{2} \equiv E_2(t,\xi)  \ .
\end{align}

 \begin{figure}[t]
    \centering
    \includegraphics[width=0.48\textwidth]{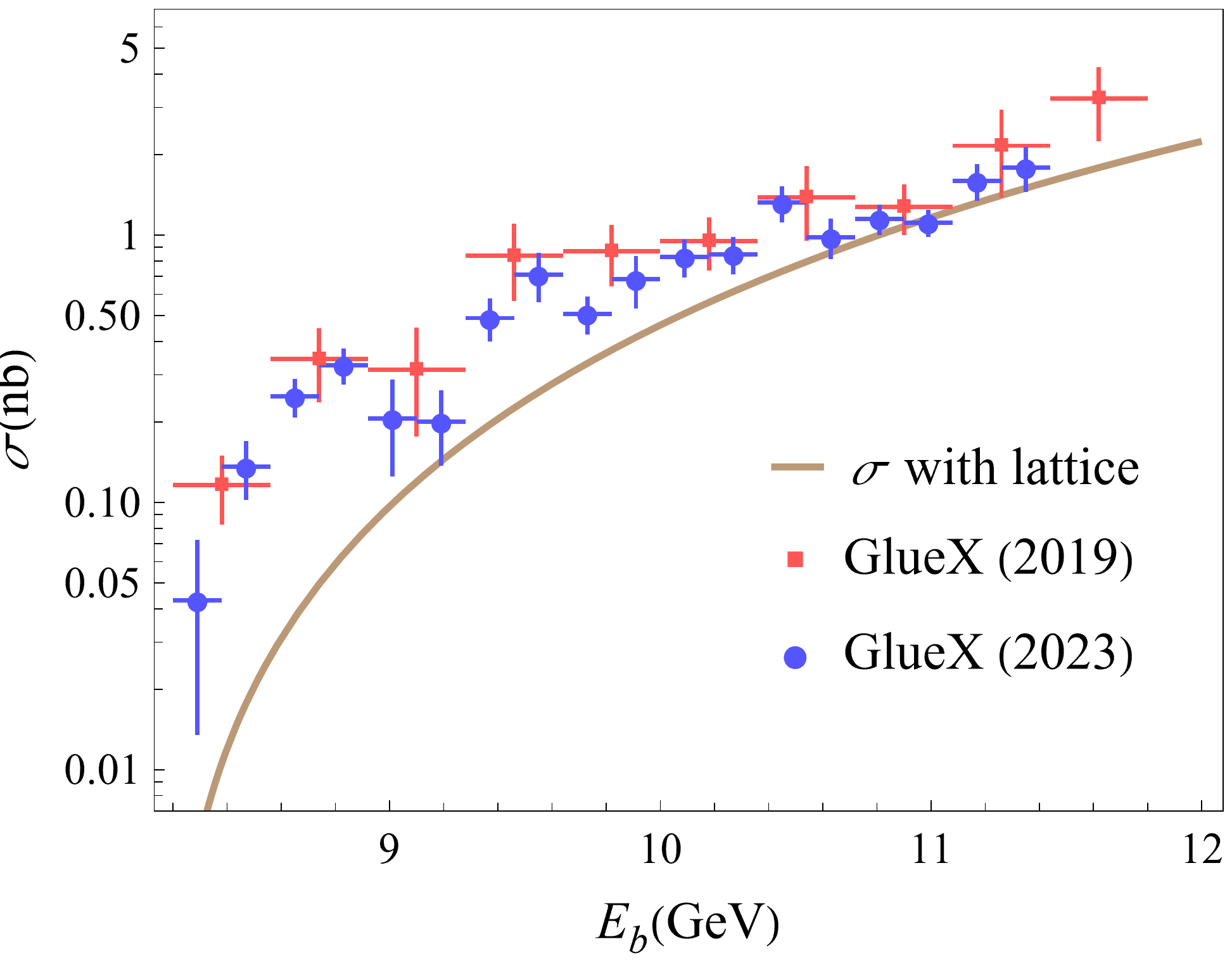}
    \caption
    {\small A comparison of the total cross-sections predicted with gluonic GFFs from the improved lattice simulation~\cite{Pefkou:2021fni} to the two measurements by the GlueX collaboration in 2019~\cite{GlueX:2019mkq} and 2023~\cite{Adhikari:2023fcr}. The $J/\psi$ 007 experiment~\cite{Duran:2022xag} did not measure the total cross-section directly and will be discussed in the next section for the differential cross-section analysis.}
    \label{fig:rawxsec}
\end{figure}

In FIG. \ref{fig:rawxsec}, we compare the total cross-sections predicted with the improved lattice simulation of the gluonic GFFs~\cite{Pefkou:2021fni} to the two measurements by the GlueX collaboration in 2019~\cite{GlueX:2019mkq} and 2023~\cite{Adhikari:2023fcr} respectively without tuning any parameters. The consistency seems better compared to the one in previous work before the revision~\cite{Guo:2021ibg}. More detailed analyses regarding the extraction of GFFs with differential cross-section measurements by both the GlueX collaboration and the $J/\psi$ 007 experiment~\cite{Duran:2022xag} will be presented in the next section.

\section{ Analyses of the near-threshold \texorpdfstring{$J/\psi$}{JPSI} production data}

To extract the gluonic GFFs from the near-threshold $J/\psi$ photoproduction measurements, recall that the cross-section formula in the previous work reads~\cite{Guo:2021ibg}:
\begin{equation}
\label{xsec}
\begin{split}
\frac{d \sigma}{d t}= \frac{ \alpha_{\rm EM}e_Q^2}{ 4\left(W^2-M_N^2\right)^2}\frac{ (16\pi\alpha_S)^2}{3M_V^3}|\psi_{\rm NR}(0)|^2 |G(t,\xi)|^2\ .
\end{split}
\end{equation}
The same formula applies after the revision except that the revised hadronic matrix element $G(t,\xi)$ in eq. (\ref{gt2ff}) has an extra factor of $4$.

We parameterize the two GFFs $A_g(t)$ and $C_g(t)$ in tripole forms:
\begin{align}\label{eq:tripoleA}
A_g(t)&=\frac{A_g(0)}{\left(1-\frac{t}{m_A^2}\right)^3}\ ,\\
\label{eq:tripoleC}
C_g(t)&=\frac{C_g(0)}{\left(1-\frac{t}{m_C^2}\right)^3}\ ,
\end{align}
ignoring the $B_g(t)$. The forward $A_g(0)$ is fixed according to the gluon PDF from global analysis to be 0.414~\cite{Hou:2019efy}. Then we are left with three parameters: $m_A$, $C_g(0)$, and $m_C$ to be determined from the near-threshold production measurements, for which we consider the combination of the recently published data from the $J/\psi$ 007 experiment~\cite{Duran:2022xag} and the GlueX collaboration~\cite{Adhikari:2023fcr}.

\begin{figure}[t]
    \centering
    \includegraphics[width=0.48\textwidth]{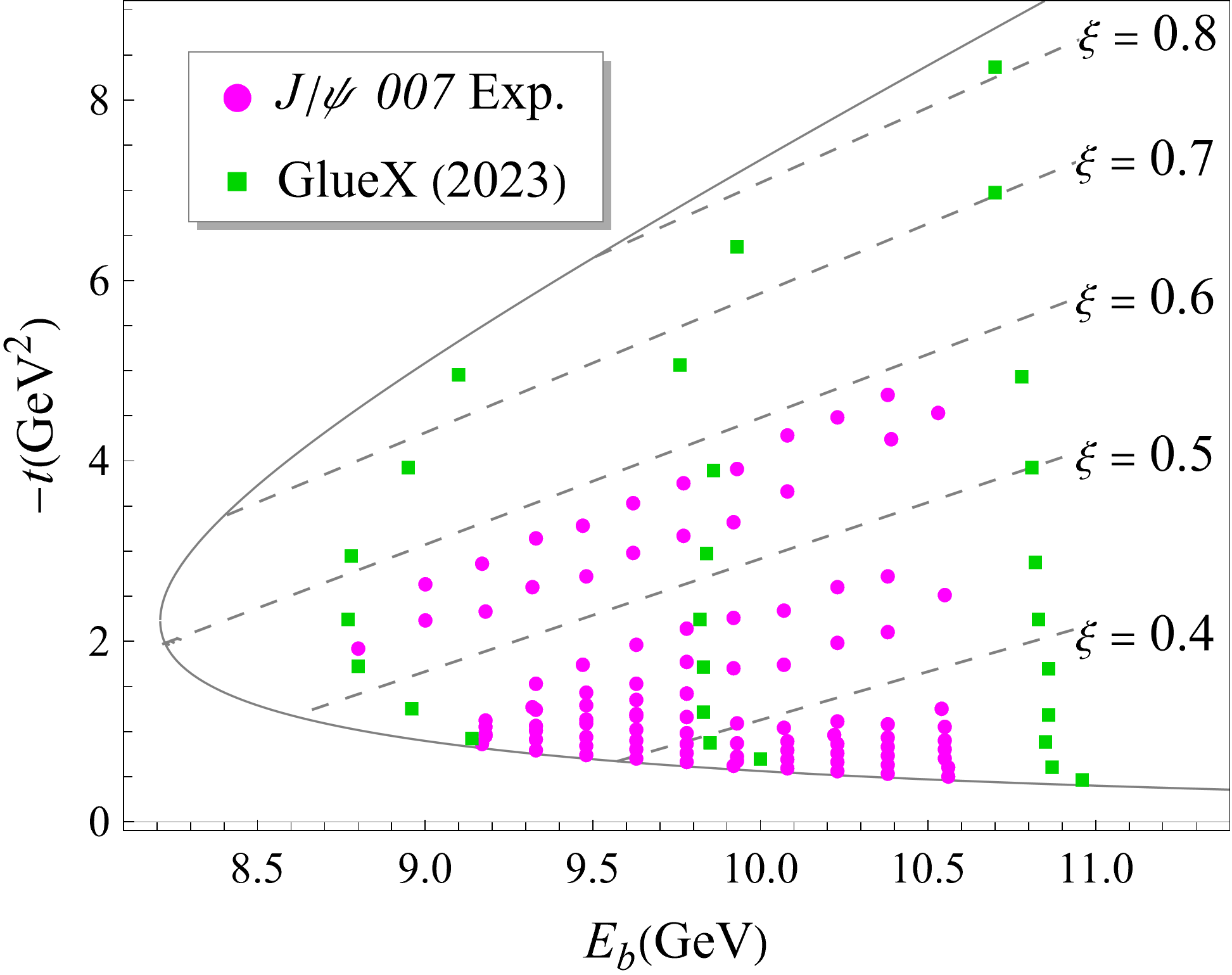}
    \caption
    {\small Differential cross-section data points together with contours of equal $\xi$ on the $(E_b,-t)$ plane in the kinematically allowed region with $M_{J/\psi}=3.097$ GeV. Each dot represents a data point from the $J/\psi$ 007 experiment at JLab Hall C (circle)~\cite{Duran:2022xag} or GlueX collaboration at Hall D (square)~\cite{Adhikari:2023fcr}.}
    \label{fig:jpsixiplot}
\end{figure}

Before moving on to the data analysis, we must first comment that the relation between the near-threshold $J/\psi$ production cross-sections and the gluonic GFFs is justified in the heavy quark limit only, where the momentum transfer squared $|t|$ approaches infinity and the skewness parameter $\xi$ approaches 1. Thus, the extraction of the gluonic GFFs relies on the so-called large $\xi$ expansion that is most applicable in the $\xi \to 1$ limit. However, we can only have measurements with finite momentum transfer squared $|t|$ and skewness $\xi<1$ in reality ---  the $J/\psi$ 007 measurements only cover the region with $\xi<0.6$ whereas the GlueX measurements extend to larger-$\xi$ region but with limited data points as shown in FIG. \ref{fig:jpsixiplot}. Out of the 124 differential cross-section data points combining the $J/\psi$ 007 and GlueX measurements, 85 have $\xi>0.4$ while only 33 have $\xi>0.5$. Also associated with the increasing $\xi$ is the decreasing quality in the data due to the lack of events at large $|t|$, putting additional challenge to the reliable extraction of the gluonic GFFs.

Accordingly, we can either focus on the large-$|t|$ data where the statistical uncertainties will be huge, or consider the medium or even lower-$|t|$ data as well where the systematical uncertainties could be the dominant effect. These possibilities will be studied in more details in the following subsections.

\subsection{Analysis with large-\texorpdfstring{$|t|$}{t} data}

We start with a rather extreme option where we  consider the data with large momentum transfer squared $|t|$ only, i.e., we select the data with $\xi>0.5$\footnote{The large $|t|$ and large $\xi$ conditions can be used interchangeably near the threshold as shown in FIG. \ref{fig:jpsixiplot} --- it requires a minimum $|t|$ to reach a certain $\xi$ and larger $\xi$ requires larger $|t|$.}. As mentioned above, there are only 33 data points in this region, which are too few to determine all three parameters $m_A$, $C_g(0)$, and $m_C$.
Indeed, by fitting all three of them to the data with $\xi>0.5$ by a standard $\chi^2$ analysis with the \texttt{iminuit} Python interface of \texttt{Minuit2} package~\cite{iminuit,James:1975dr}, we obtain $m^{\xi>0.5}_A=2.38\pm0.08\text{ GeV}$, $C^{\xi>0.5}_g(0)=-16\pm34$ and $m^{\xi>0.5}_C=0.60\pm0.26\text{ GeV}$. The unreasonably large best-fit value and statistical uncertainty of $C_g(0)$ indicate that the $C_g(t)$ form factor cannot be effectively constrained with the large-$|t|$ data. The $A_g(t)$ form factor, on the other hand, is better constrained benefiting from the forward constraint from gluon PDFs. Consequently, we have to utilize other information about the gluonic GFFs, e.g., the ones from lattice simulations~\cite{Shanahan:2018pib,Pefkou:2021fni}, as the reference values to fix the undetermined parameters and avoid potential overfitting.

\begin{table}[t]
    \centering
    \def\arraystretch{1.35}
    \begin{tabular}{c |c c c c c}
    \hline \hline
    ~Set~ & ~$A_g(0)$~  & ~$m_A (\text{GeV}) $~ & ~$C_g(0)$~  & ~$m_C (\text{GeV}) $ ~& ~red. $\chi^2$~\\
    \hline
    ref. & 0.414 & 1.64 & -0.48 &1.07 & - \\
    1 & fixed & 2.38(08) & -16(34) &0.60(26)&1.34 \\
    2 &  fixed & 2.53(12) & fixed & 1.41(09)&1.39\\
    3 &  fixed & 2.46(08) & -1.20(20)& fixed&1.33 \\
    4 & fixed & fixed & 0.014(03) & 13(14)&1.25 \\
    5 & fixed & 2.14(03) & fixed & fixed&1.70 \\
    6 & fixed & fixed & 0.29(06) & fixed&2.63 \\
    \hline\hline
    \end{tabular}
    \caption{A summary of the best-fit parameters of the $\xi>0.5$ data with reduced $\chi^2$ fitting to the combined differential cross-section data from the $J/\psi$ 007 experiment~\cite{Duran:2022xag} and the GlueX collaboration~\cite{Adhikari:2023fcr}. The reference values are from the gluon PDF~\cite{Hou:2019efy} for the $A_g(0)$ and lattice simulations~\cite{Pefkou:2021fni} for the other three. Parameters listed as ``fixed" are fixed to be the reference values. }
    \label{table:sumxi05}
\end{table}

In TABLE \ref{table:sumxi05}, we summarize the results from various fits that fix the parameters differently, where in the first row we list the reference values from the global-fitted gluon PDF for the $A_g(0)$~\cite{Hou:2019efy} and lattice simulations of GFFs~\cite{Pefkou:2021fni} for the other three parameters. Among all the fits, set 2 and 3 seem more realistic that exhibit no signs of overfitting and describe the data fine with reduced $\chi^2$s around 1.3. On the other hand, both set 1 and 4 have undetermined parameters with unreasonably large values and uncertainties though their reduced $\chi^2$s look fine. Such observation indicates the potential overfitting in these sets. As for set 5 and 6, their large reduced $\chi^2$s imply that a one-parameter fit may not be able to describe the large-$|t|$ data well.

We also comment on the size of reduced $\chi^2$s in the fits which seems rather large for fits with only 33 data points. The main reason is the anomalously rising behavior in the $t$-dependence of the measured differential cross-sections by the GlueX collaboration, which has been discussed with more details in ref.~\cite{Adhikari:2023fcr} (see for instance the FIG. 13 therein). Such behaviors were not observed in the $J/\psi$ 007 experiment, partially due to its limit kinematical coverage at large $|t|$ as shown in FIG. \ref{fig:jpsixiplot}. Other than that, the GlueX measurements are in good agreement with the measurements by the $J/\psi$ 007 experiment. Also note that the GlueX data have about $20\%$ normalization  uncertainties (about $4\%$ for the $J/\psi$ 007 experiment) that we did not include since the two data sets seem consistent except for this rising $t$-dependence observed by the GlueX. This would lead to lower reduced $\chi^2$ if included.

\begin{figure}[t]
    \centering
    \includegraphics[width=0.48\textwidth]{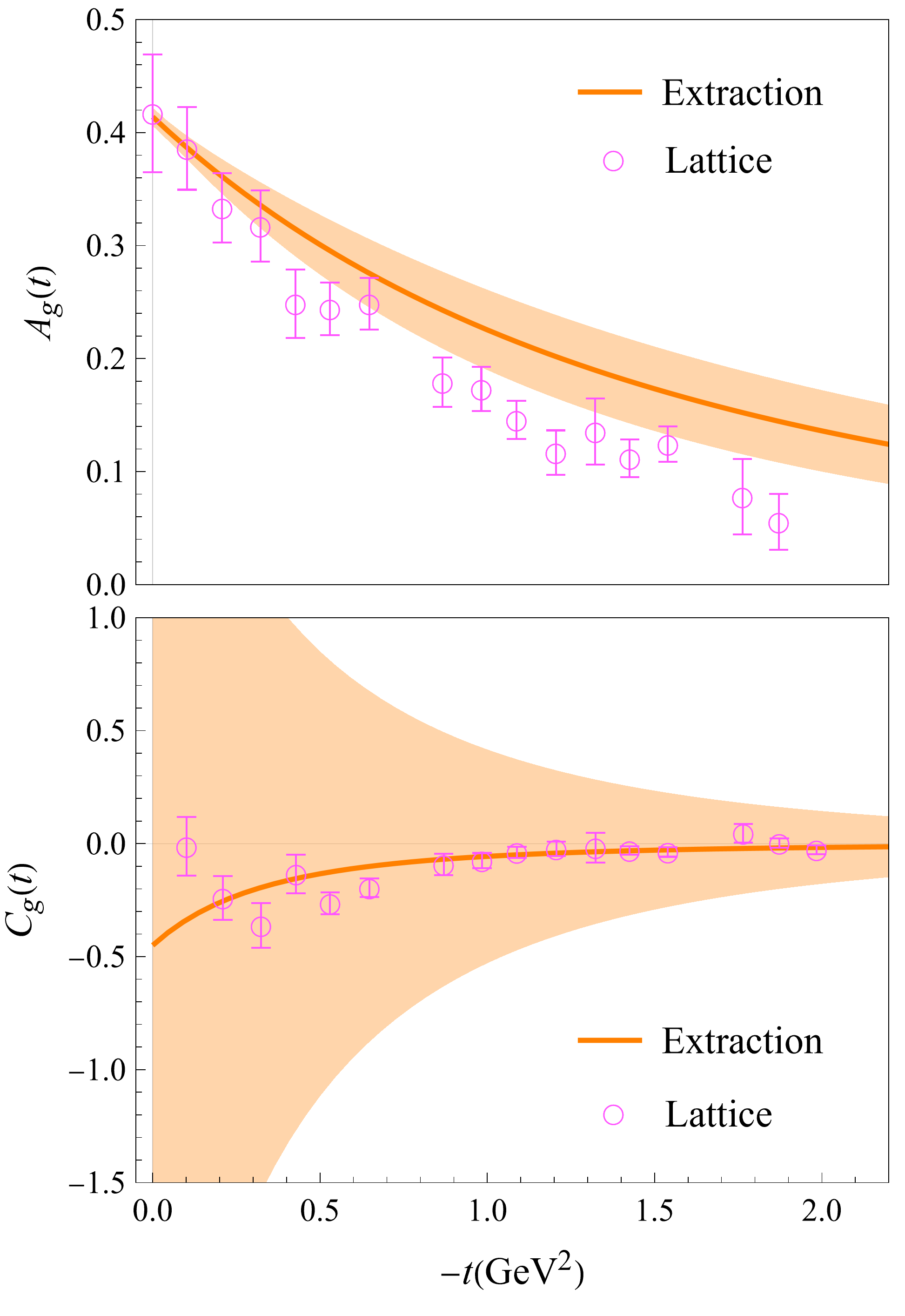}
    \caption
    {\small A comparison of the $A_g(t)$ (up) and $C_g(t)$ (down) form factor extracted with the differential cross-section data from the $J/\psi$ 007 experiment only~\cite{Duran:2022xag} selecting $\xi>0.5$ and the lattice simulation results~\cite{Pefkou:2021fni}. Bands correspond to $1\sigma$ statistical uncertainties.}
    \label{fig:fffitxicut}
\end{figure}

For the analysis here, when we select large-$|t|$ data, such behaviors observed in the large-$|t|$ region will be more pronounced, which cannot be described well by the tripole form used here.  This caused the overfitting behavior observed for the $C_g(0)$ here, and it will continue to make the reduced $\chi^2$s larger even when we extend to include the medium or lower-$|t|$ data as we will show in the following subsections. To check this argument, we perform the same fits but to the $J/\psi$ 007 data only, and obtain much lower reduced $\chi^2$s for the fits here and in the following subsections. For instance, we obtain $m^{\xi>0.5}_A=2.10\pm0.29\text{ GeV}$, $C^{\xi>0.5}_g(0)=-0.5\pm 2.6$ and $m^{\xi>0.5}_C=1\pm 1.6\text{ GeV}$ with a reduced $\chi^2$ of 0.46 when fitting to only the $J/\psi$ 007 data with $\xi>0.5$. Although the relative uncertainties are larger due to the reduced sample size, the overfitting behavior seems to disappear. In FIG. \ref{fig:fffitxicut}, we also present the corresponding extracted GFFs and compare them with the lattice results. We note that the rising $t$-dependence of the measured differential cross-section by the GlueX collaboration might be due to other sources beyond the scope of this work.

With these results, we remark that with the limited data at large $|t|$, not much information can be extracted about the gluonic GFFs, especially the $C_g(t)$ form factor. Excluding the fits with undetermined parameters or large reduced $\chi^2$s, the remaining ones cannot constrain the $C_g(t)$ form factor well unless when one of the parameters is fixed. While the $A_g(t)$ form factor is better constrained owing to the forward constraint from the gluon PDF, the extracted tripole mass $m_A$ still depends on the values of the fixed parameters, the $C_g(0)$ particularly. Since the lattice simulation of the $C_g(t)$ form factor has the largest uncertainties at $t=0$~\cite{Pefkou:2021fni}, the reliability of the extracted $m_A$ would be affected accordingly. Therefore, it is crucial to obtain more data with higher quality in the analysis to better constrain the gluonic GFFs.

\subsection{Analysis including medium-\texorpdfstring{$|t|$}{t} data}

Given the limited constraining power with only the large-$|t|$ data due to the insufficient data, one compromising choice is to include more data with medium $|t|$. As we decrease the cut in $|t|$ or the skewness $\xi$, we effectively reduce the statistical uncertainties while enhancing the systematical uncertainties. Then, the general philosophy is to find the cut that balances the two uncertainties to maximally utilize the data. Although the proper estimation of the systematical uncertainties from the higher order effects and large $\xi$ expansion will be extremely involving and beyond the scope of this work, we could still include the medium-$|t|$ data in the fit and study their effects for discussion.

\begin{table}[t]
    \centering
    \def\arraystretch{1.35}
    \begin{tabular}{{c |c c c c c}}
    \hline \hline
    ~Set~ & ~$A_g(0)$~  & ~$m_A (\text{GeV}) $~ & ~$C_g(0)$~  & ~$m_C (\text{GeV}) $ ~& ~red. $\chi^2$~\\
    \hline
    ref. & 0.414 & 1.64 & -0.48 &1.07 & - \\
    1 & fixed & 2.07(05) & -1.21(37) &0.91(10) & 1.42 \\
    2 &  fixed & 2.25(05) & fixed & 1.38(04)& 1.64\\
    3 &  fixed & 2.14(03) & -0.81(03)& fixed& 1.43 \\
    4 & fixed & 1.88(01) & fixed & fixed& 2.38 \\
    5 & fixed & fixed & -0.29(02) & fixed& 3.74 \\
    \hline\hline
    \end{tabular}
    \caption{A summary of the best-fit parameters of the $\xi>0.4$ data with reduced $\chi^2$ fitting to the combined differential cross-section data from the $J/\psi$ 007 experiment~\cite{Duran:2022xag} and the GlueX collaboration~\cite{Adhikari:2023fcr}. The reference values are from the gluon PDF~\cite{Hou:2019efy} for the $A_g(0)$ and lattice simulations~\cite{Pefkou:2021fni} for the other three. Parameters listed as ``fixed" are fixed to be the reference values.}
    \label{table:sumxi04}
\end{table}

With that in mind, we repeat the above analysis with the 85 data when selecting $\xi>0.4$. The extended set of data does constrain the three parameters better. Consequently, we obtain $m^{\xi>0.4}_A=2.07\pm0.05\text{ GeV}$, $C^{\xi>0.4}_g(0)=-1.2\pm0.4$, and $m^{\xi>0.4}_C=0.91\pm0.10\text{ GeV}$ from a three-parameter fit. This, together with the other fits that fix the parameters differently, is summarized in TABLE \ref{table:sumxi04} as the set 1. Unlike the previous case where many of the parameters cannot be determined, no signs of overfitting are observed here by virtue of the extra inputs. Among all the fits, set 1 has the lowest reduced $\chi^2$ as expected. Meanwhile, since the $m_C$ obtained in set 1 is close to the reference value, the two-parameter fit in set 3 that fixes the $m_C$ looks similar. The other two-parameter fit in set 2 shows slightly worse results, whereas the one-parameter fits in set 4 and 5 do not seem to work. Thus, we consider the three-parameter fit in set 1 for further analysis.

\begin{figure}[t]
    \centering
    \includegraphics[width=0.48\textwidth]{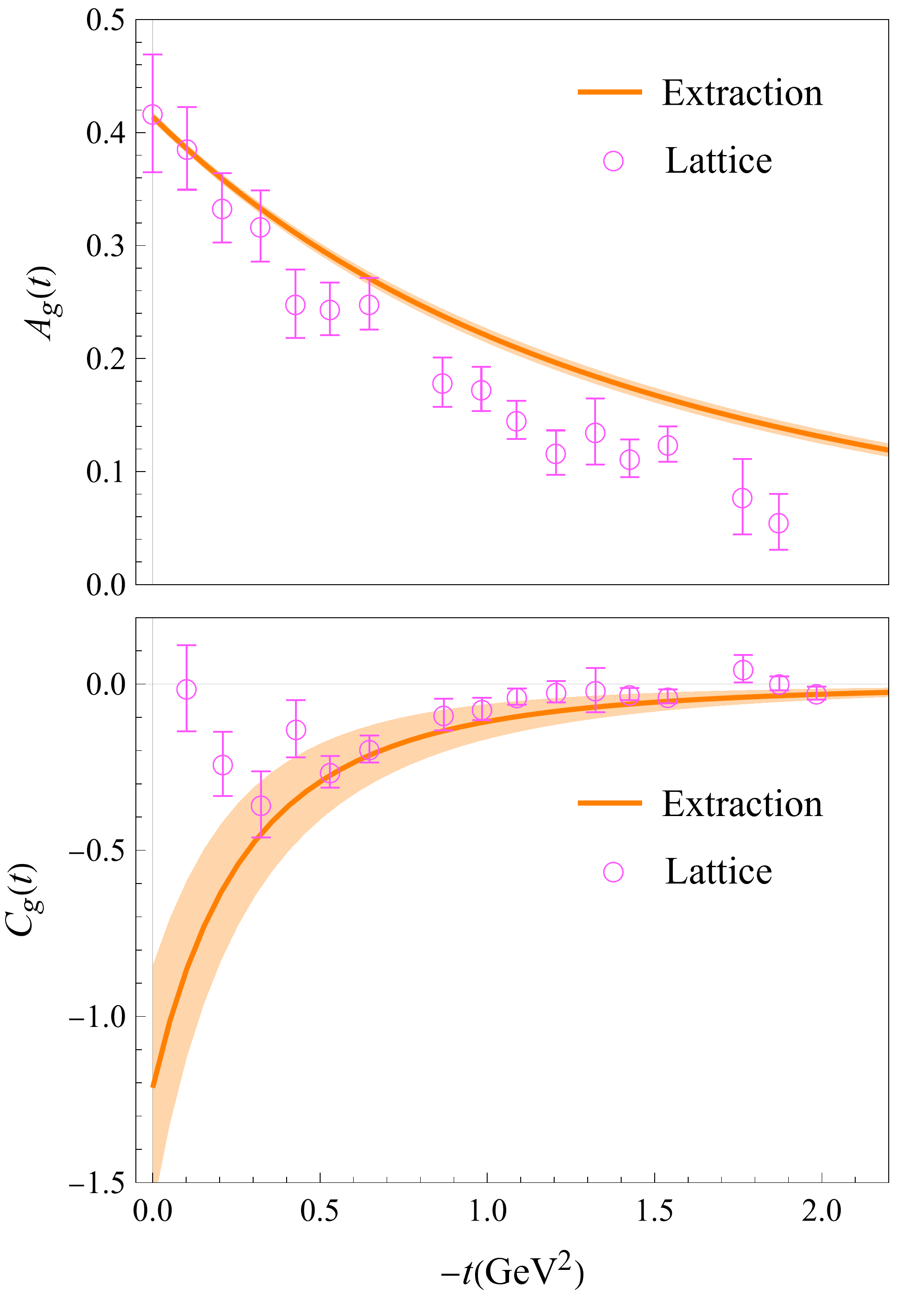}
    \caption
    {\small A comparison of the $A_g(t)$ (up) and $C_g(t)$ (down) form factor extracted with the combined differential cross-section data from the $J/\psi$ 007 experiment~\cite{Duran:2022xag} and the GlueX collaboration~\cite{Adhikari:2023fcr} selecting $\xi>0.4$ and the lattice simulation results~\cite{Pefkou:2021fni}. Bands correspond to $1\sigma$ statistical uncertainties.}
    \label{fig:fffitxicut2}
\end{figure}

In FIG. \ref{fig:fffitxicut2}, we compare the GFFs extracted in set 1 with the ones from lattice simulation~\cite{Pefkou:2021fni}. Still, a negative $C_g(0)$ is favored when including the medium-$|t|$ data which is consistent with the lattice simulation. The uncertainty associated with the extracted $C_g(0)$ has been improved though still sizable, whereas the $A_g(t)$ form factor remains to be better constrained due to the forward constraint from the gluon PDF. Thus, we note that when including the medium-$|t|$ data the observation is consistent with the large-$|t|$ extraction. However, the gluonic GFFs are better constrained with the extra data, allowing us to roughly extract the gluon GFFs. Keeping in mind that there are still model-dependence and systematical uncertainties to be clarified in the extraction, we shall note that, as mentioned above, more data with higher quality at large/medium $|t|$ are crucial for better determination of the gluonic GFFs in this framework.

 \begin{figure}[t]
    \centering
    \includegraphics[width=0.48\textwidth]{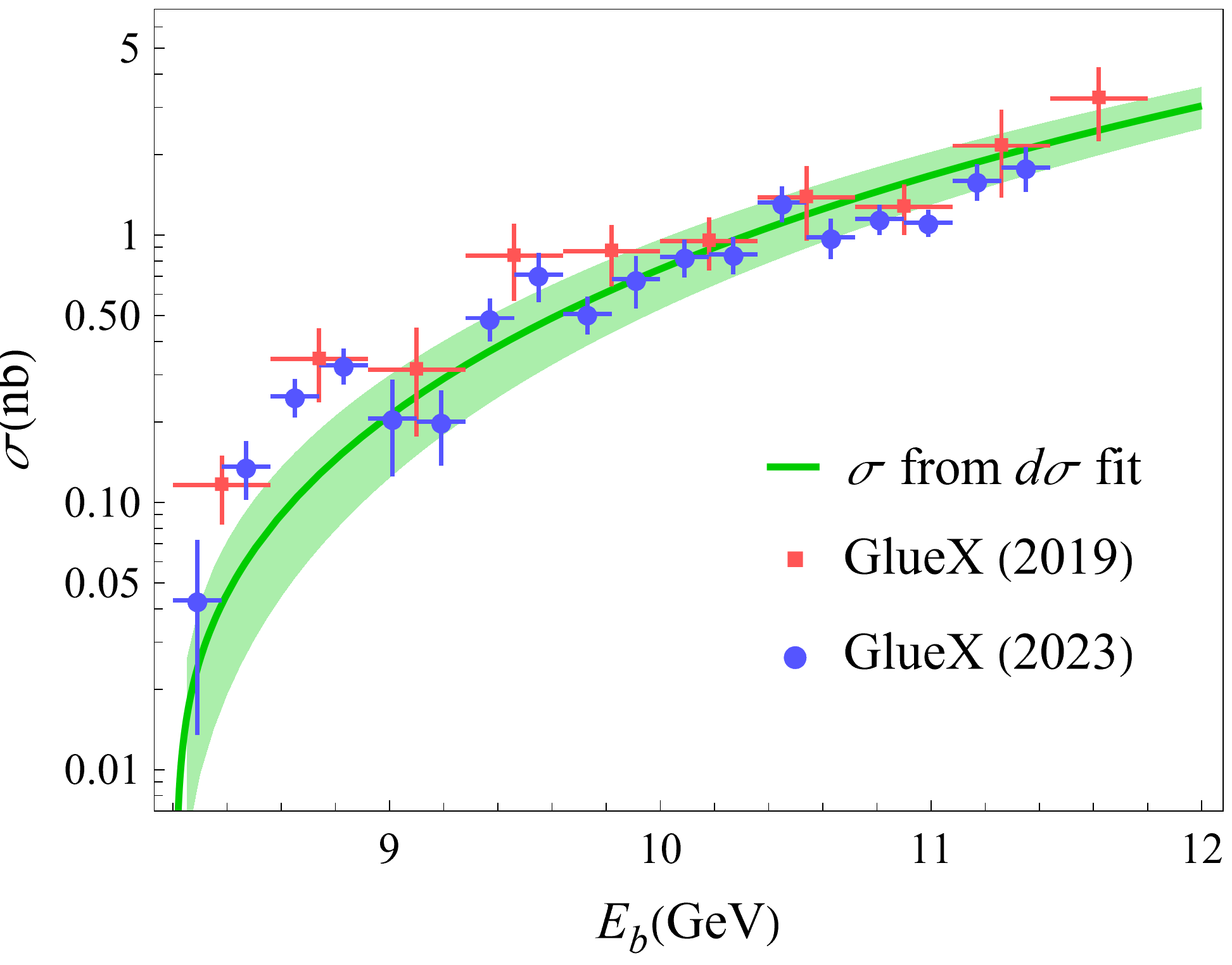}
    \caption
    {\small A comparison of the total cross-sections calculated based on the fit to the combined differential cross-section data from the $J/\psi$ 007 experiment~\cite{Duran:2022xag} and the GlueX collaboration~\cite{Adhikari:2023fcr} selecting $\xi>0.4$ and the two GlueX measurements~\cite{GlueX:2019mkq,Adhikari:2023fcr}. The band corresponds to $1\sigma$ statistical uncertainties.}
    \label{fig:fitxsec}
\end{figure}

In FIG. \ref{fig:fitxsec}, we also compare the total cross-section calculated based on the fit in set 1 to the two measurements by the GlueX collaboration in 2019~\cite{GlueX:2019mkq} and 2023~\cite{Adhikari:2023fcr}. Although these total cross-section data are not considered in the fit, they still show good agreement.

\subsection{Analysis with full data}

In the last subsection, we also present the fit to all the 124 differential cross-section data. Since about 1/3 of these data have $\xi<0.4$ and they will be weighted more than the others in the fit resulting from their lower relative uncertainties, the systematical uncertainties in this case can get out of control. Therefore, we consider it as an exercise to illustrate the effect of the lower-$|t|$ data in the analysis. It has to be kept in mind that these results should NOT be taken too seriously due to the potentially large systematical uncertainties associated.

 \begin{figure}[t]
    \centering
    \includegraphics[width=0.48\textwidth]{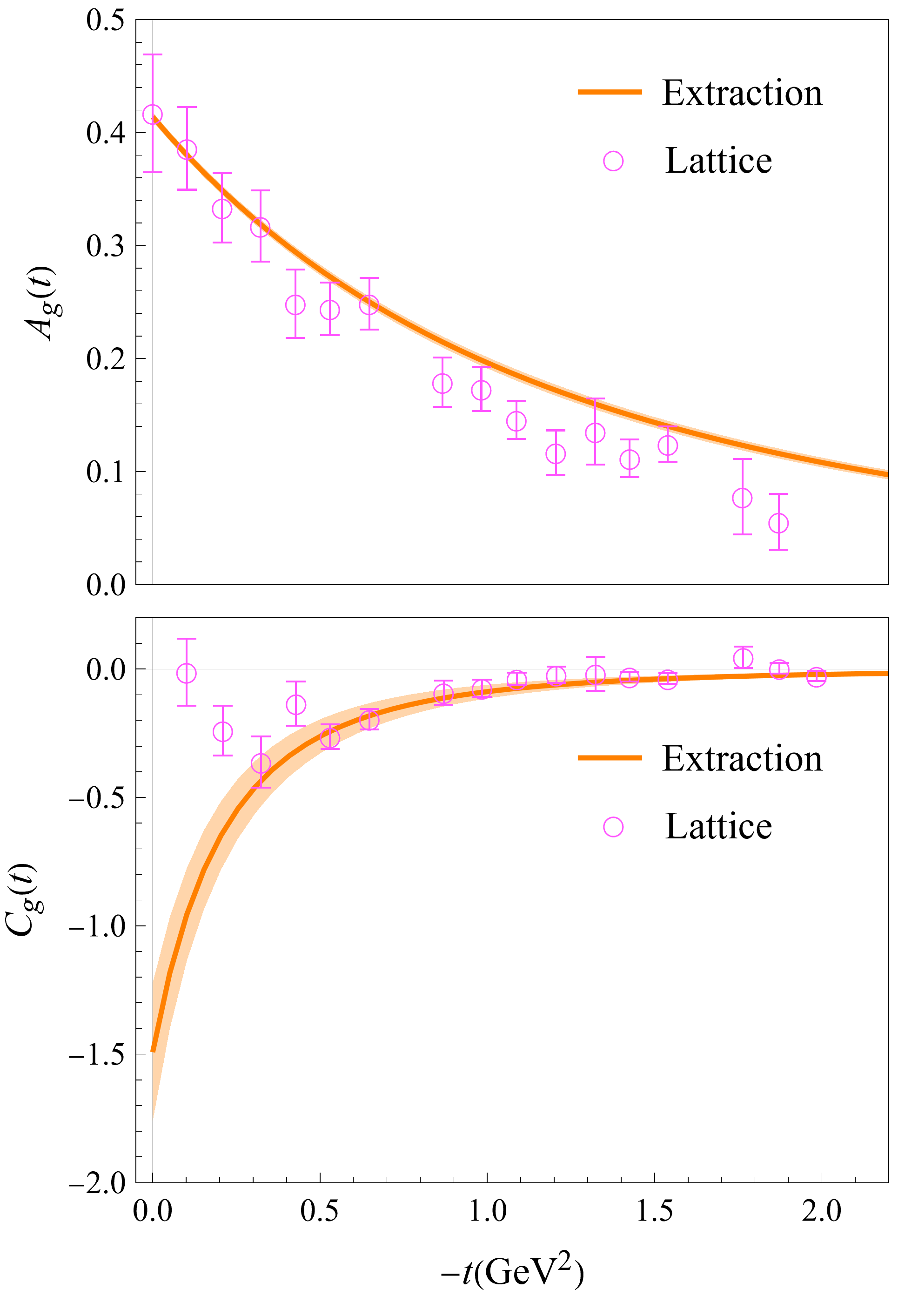}
    \caption
    {\small A comparison of the $A_g(t)$ (up) and $C_g(t)$ (down) form factor extracted with all the combined differential cross-section data from the $J/\psi$ 007 experiment~\cite{Duran:2022xag} and the GlueX collaboration~\cite{Adhikari:2023fcr} and the lattice simulation results~\cite{Pefkou:2021fni}. Bands correspond to $1\sigma$ statistical uncertainties.}
    \label{fig:fffit}
\end{figure}

Being cautious of that, we fit the three parameters $m_A$, $C_g(0)$, and $m_C$ to all the data, and obtain $m_A=1.88\pm0.03\text{ GeV}$, $C_g(0)=-1.49\pm0.27$, and $m_C=0.8\pm0.05\text{ GeV}$ with the reduced $\chi^2$ as 1.89. An explicit comparison of the extracted GFFs with the lattice simulation~\cite{Pefkou:2021fni} is shown in FIG. \ref{fig:fffit}. Quite surprisingly, the extracted GFFs agree well with the lattice simulation, which is in accord with the previous finding in FIG. \ref{fig:rawxsec} that the revised cross-section formula produces nice consistency between the lattice simulations and the near-threshold $J/\psi$ production cross-sections. However, since the GFFs from lattice simulation are mostly in the small-$|t|$ region where the systematical uncertainty could be large, the reliability of these statements might be challenged by that accordingly.

\section{Extracted mass and scalar radii}

To make the comparison more intuitive, in this section we also consider the extracted mass and scalar radii of the proton. They
can be defined in terms of the GFFs as~\cite{Ji:2021mtz},
\begin{equation}\label{eq:radiidef}
\begin{aligned}
\left\langle r^{2}_m\right\rangle &=\left[6 \left.\frac{d A\left(t\right)}{d t}\right|_{t=0}-6 ~\frac{C(0)}{M_N^{2}}\right]\ ,\\
\left\langle r^{2}_s\right\rangle &=\left[6 \left.\frac{d A\left(t\right)}{d t}\right|_{t=0}-18 ~\frac{C(0)}{M_N^2}\right]\ ,
\end{aligned}
\end{equation}
where the $A(t)$ and $C(t)$ form factors are the sums of the quark and gluon GFFs, respectively. We note that one complexity arises from the $\bar C(t)$ form factor when individual contributions from the quark and gluon are considered~\cite{Ji:2021mtz}. The $\bar C(t)$ terms exist in the matrix elements of the energy momentum tensor (EMT) of quark and gluon separately, while they cancel each other when summing over the quark and gluon: $\bar C_q(t) + \bar C_g(t) =0$ due to current conservation~\cite{Ji:1996ek}. Furthermore, the separation of the quark and gluon GFFs depends on the renormalization scheme and scale. Therefore, proton mass and scalar radii can only be obtained unambiguously with both the quark and gluon GFFs. Besides the gluon GFFs extracted here, the quark GPDs and GFFs can be probed by processes such as deeply virtual Compton scattering (DVCS)~\cite{Ji:1996nm} where the $C_q(t)$ form factor can be extracted with dispersive analysis~\cite{Burkert:2018bqq, Kumericki:2019ddg}. In addition, the lattice QCD simulation also plays a critical role in obtaining the quark GPDs and GFFs from first-principle calculations~\cite{Hagler:2009ni,Bali:2018zgl,Alexandrou:2019ali,Alexandrou:2020zbe,Constantinou:2020hdm,Lin:2020rxa,Lin:2021brq,Bhattacharya:2023ays}.

With that in mind, we will consider the extra inputs for the quark GFFs as well to compare the proton mass and scalar radii. It is quite apparent with eq. (\ref{eq:radiidef}) that these radii depend on the first derivative of the $A(t)$ form factor and the value of  the $C(t)$ form factor at $t=0$, but not the derivative of $C(t)$. Thus, with the tripole form parameterization in eqs. (\ref{eq:tripoleA}) and (\ref{eq:tripoleC}), we will need $A_q(0)$, $C_q(0)$ and the tripole mass $m_{A,q}$ for the quark GFFs. We will ignore the contributions from the $\bar C(t)$ form factors since they do not affect the full radii when combing quarks and gluons. Besides, they are of higher twist and much harder to obtain.

The $A_q(0)$ can be simply taken from the global quark PDFs to be $A_{u+d}(0)=0.543\pm 0.007$~\cite{Hou:2019efy}, where we ignore the contributions from strange and heavier quarks. The $m_{A,q}$ and  $C_q(0)$ cannot be obtained directly from forward measurements. For the $C_q(0)$, we use the $C_{u+d}(0)$ from the dispersive analysis of the DVCS measurements to be $-0.41\pm0.12$ where the $30\%$ relative uncertainty is estimated based on the potential contamination from higher moments~\cite{Burkert:2018bqq,Burkert:2023wzr}. As for the $m_{A,q}$, since it is not well constrained by the experiments, we consider the dipole mass from the recent lattice results~\cite{Bhattacharya:2023ays}. We obtain $m_{A,u+d}^{\rm{lat, dipole}}=1.70\pm0.06$ based on a dipole fit to the $A_{20}^{u+d}(t)$ therein\footnote{Note that a dipole rather than tripole form was used to fit the lattice results, of which the difference will be taken care of.}.

\begin{figure}[t]
	\centering
	\includegraphics[width=0.48\textwidth]{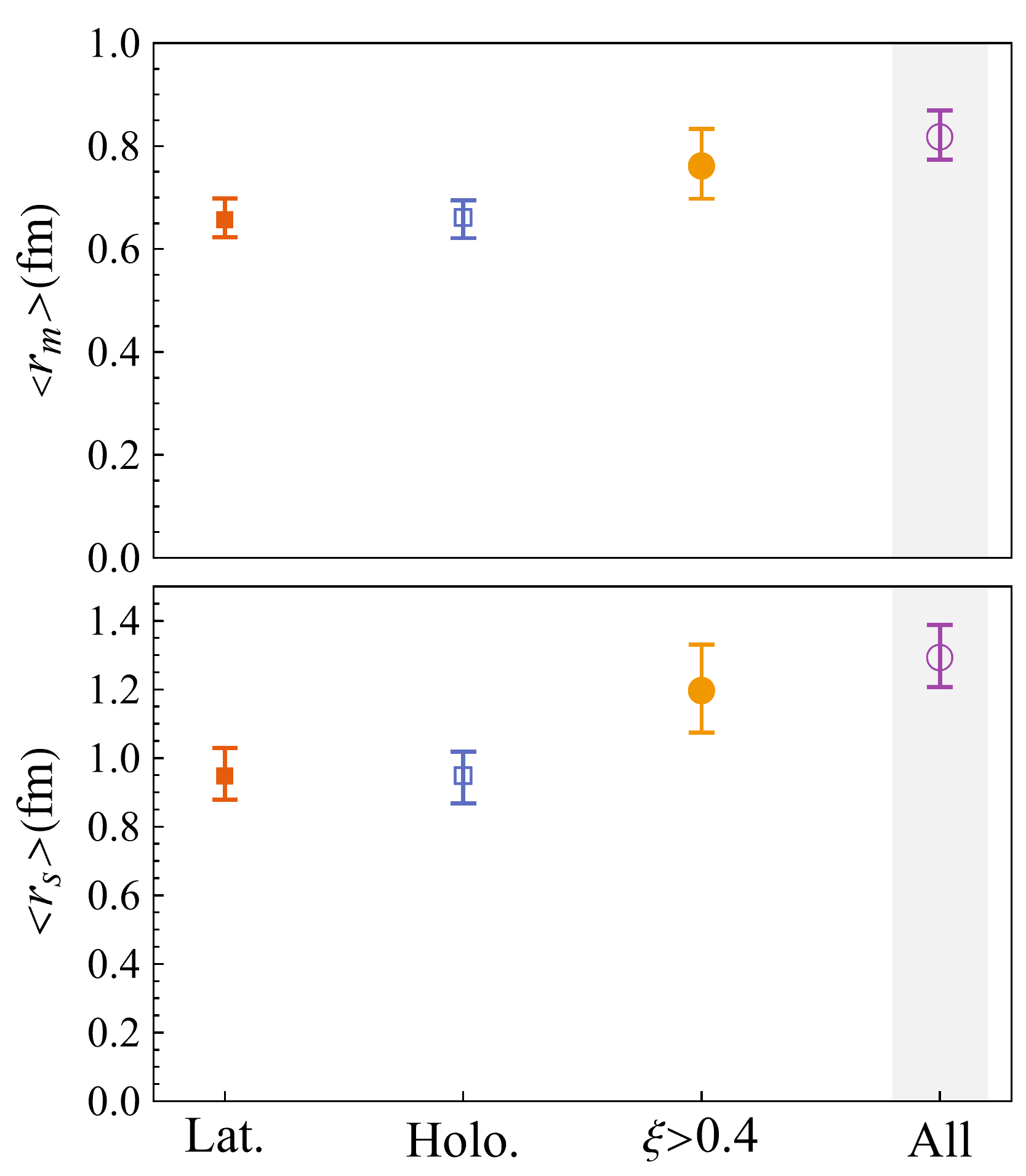}
	\caption
	{\small Comparisons of the extracted proton mass (up) and scalar (down) radii with the same quark GFFs but different gluon GFFs. The four values from left to right take the gluon GFFs from lattice simulation~\cite{Pefkou:2021fni}, holographic QCD extraction~\cite{Mamo:2022eui} with the $J/\psi$ 007 measurements~\cite{Duran:2022xag}, the extraction in this work with $\xi>0.4$ data combining the $J/\psi$ 007 experiment and the GlueX measurements~\cite{Adhikari:2023fcr}, and the extraction with all data combining the $J/\psi$ 007 experiment and the GlueX measurements, respectively. The gray bands indicate potentially large systematical uncertainties besides the statistical ones.}
	\label{fig:rplot}
\end{figure}

Combing these extra inputs for the quark GFFs with the gluon ones, we compare the corresponding proton mass and scalar radii in FIG. \ref{fig:rplot}. The scale/scheme dependence and evolution effects which are of higher order in $\alpha_S$ will be ignored here for simplicity. With the same quark GFFs, the four mass and scalar radii are calculated with the gluon GFFs from lattice simulations~\cite{Pefkou:2021fni}, the extraction based on holographic QCD model with the $J/\psi$ 007 measurements~\cite{Duran:2022xag}, the extraction in this work with $\xi>0.4$ data combining the $J/\psi$ 007 experiment and the GlueX measurements~\cite{Adhikari:2023fcr}, and the extraction with all data combining the $J/\psi$ 007 experiment and the GlueX measurements, respectively. Consistent with the arguments before, as we increase the cut in $\xi$, there will be fewer measurements with larger uncertainties, leading to more uncertain mass and scalar radii correspondingly. On the other hand, the fits with lower cut in $\xi$ will be associated with larger systematical uncertainties, e.g., in the fit with all data.

These radii from various extractions are in relatively good agreements, given that they have the same quark contributions. However, one may still notice that the scalar and mass radii extracted with the holographic QCD approach appear to agree better with the lattice results with smaller statistical uncertainties compared to ours. There are two reasons that account for that. First, the holographic approach works better in the Regge limit $t\to 0$ where more data with higher quality exist. Besides, there is an undetermined normalization constant in the holographic QCD approach that is manually fixed~\cite{Mamo:2022eui}, with which the statistical/systematical uncertainties associated are not accounted. Therefore, one could consider the difference between the gluonic GFFs and the scalar/mass radii extracted with the holographic QCD approach and the ones here as an estimation of the systematical uncertainties or model dependence. The main sources of the uncertainties in the extracted radii are from the $C(0)$, as the $A(0)$ is well determined from the global analysis of PDFs~\cite{Hou:2019efy}. The $C(t)$ form factor has been assigned the pressure or shear pressure interpretation analogous to the macroscopic fluid~\cite{Polyakov:2002yz,Polyakov:2018zvc,Burkert:2018bqq}, though it is argued that it should be considered as the gravitational tensor-monopole moment according to their role
in generating static gravity nearby~\cite{Ji:2021mfb,Ji:2022exr}. Thus, better constraints on the $C(t)$ form factor, especially at $t=0$, are crucial not only to improve the quality of the extracted radii but also to obtain a more profound understanding of the fundamental mechanic properties of the nucleon.

To emphasize the gluonic contributions, we also consider the gluonic mass and scalar radii, which are the gluonic contributions to the corresponding proton radii renormalized with the gluon momentum fraction $A_g(0)$:
\begin{equation}\label{eq:gluonradiidef}
\begin{aligned}
\left\langle r^{2}_m\right\rangle_g &=\frac{1}{A_g(0)}\left[6 \left.\frac{d A_g\left(t\right)}{d t}\right|_{t=0}-6 ~\frac{C_g(0)}{M_N^{2}}\right]\ ,\\
\left\langle r^{2}_s\right\rangle_g &=\frac{1}{A_g(0)}\left[6 \left.\frac{d A_g\left(t\right)}{d t}\right|_{t=0}-18 ~\frac{C_g(0)}{M_N^2}\right]\ ,
\end{aligned}
\end{equation}
ignoring the contribution from $\bar C_g(0)$.

\begin{figure}[t]
	\centering
	\includegraphics[width=0.48\textwidth]{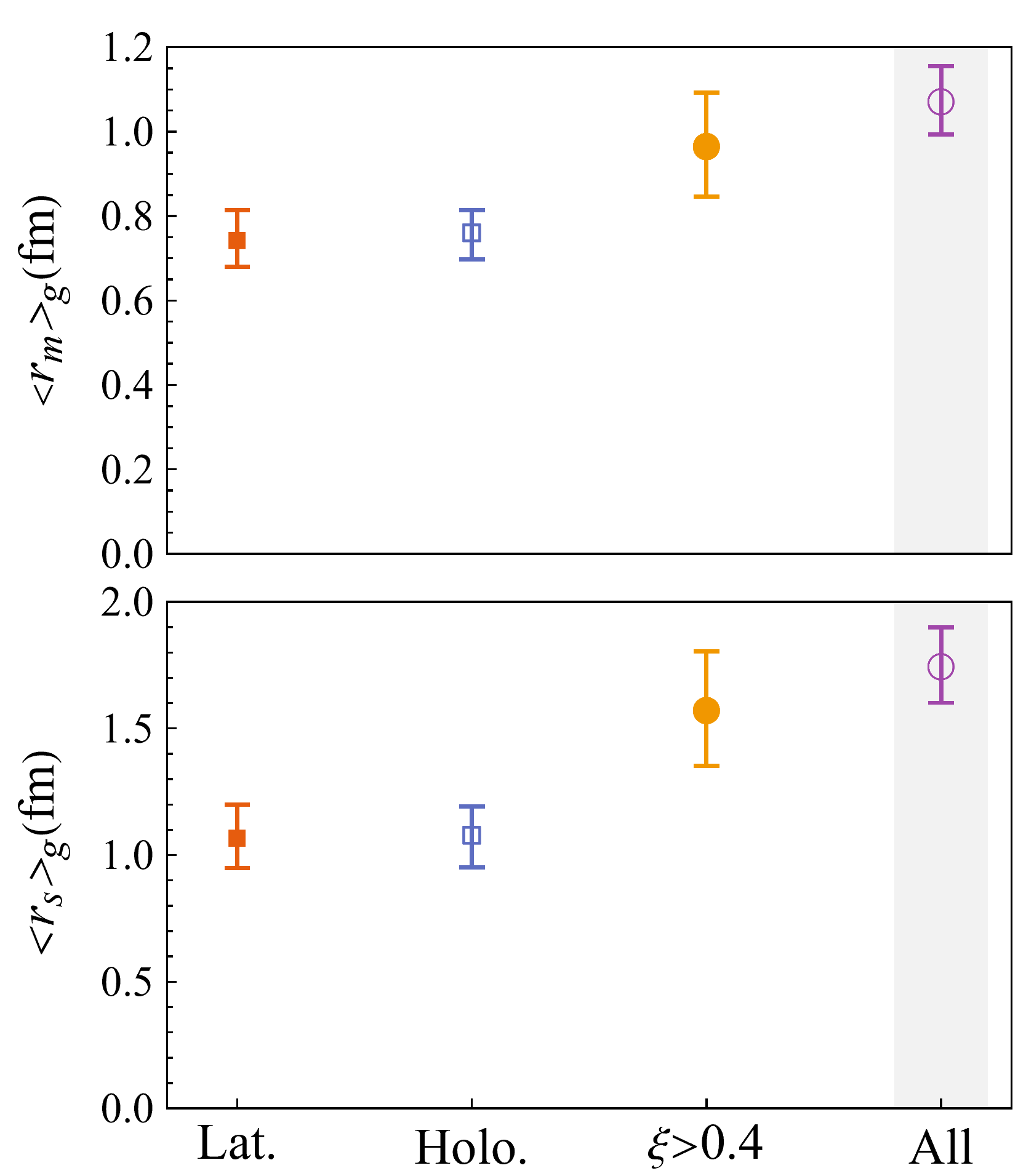}
	\caption
	{\small Comparisons of the extracted gluonic mass (up) and scalar (down) radii. The differences among the gluonic radii are more pronounced since only the gluonic contributions are considered. Note the different scales from the previous plots.}
	\label{fig:rgplot}
\end{figure}

In FIG. \ref{fig:rgplot} we compare the pure gluonic proton mass and scalar radii similar to what we did for the full mass and scalar radii. The differences among the various extractions are more significant, and accompanied by larger uncertainties correspondingly, resulting from the large deviation and uncertainties in the extracted $C_g$. Accordingly, more measurements with higher quality, especially at large $|t|$ are of vital importance to further improve the extraction of the gluonic GFFs and the mass/scalar radii from the near-threshold $J/\psi$ production measurements.

We note again that there could be comparable systematical uncertainties associated with these statistical uncertainties, which should be even more sizable for the extraction with all data here as we explained in the previous section. Therefore, we consider the extraction with $\xi>0.4$ data as the reference values of this work, which gives $\left<r_m\right>=0.77\pm0.07 \rm{~fm}$ and $\left<r_s\right>=1.20\pm0.13 \rm{~fm}$ for the full proton mass and scalar radii and $\left<r_m\right>_g=0.97\pm0.12 \rm{~fm}$ and $\left<r_s\right>_g=1.58\pm0.23 \rm{~fm}$ for the pure gluonic ones.

\section{Summary and outlook}

To summarize, in this work we revise the previous cross-section formula~\cite{Guo:2021ibg} with the factor of 2 mismatch, and perform an updated analysis with the latest lattice simulation of gluonic GFFs~\cite{Pefkou:2021fni} and the recently published data from $J/\psi$ 007 experiment~\cite{Duran:2022xag} and GlueX collaboration~\cite{Adhikari:2023fcr}. We show that with the revised formula the agreement between the gluonic GFFs from lattice simulation and extraction with GPD factorization gets improved. On the other hand, we also argue that this framework requires large momentum transfer squared $|t|$ and skewness $\xi$, and thus we perform a series of analyses to properly address their effects. We show that the gluonic GFFs can be roughly constrained with a cut of $\xi>0.4$. However, it is crucial to have more high-quality data at large $|t|$ to improve the extraction.

The critical future developments include studying the higher order corrections such as the next-to-leading order effects in the strong coupling $\alpha_S$ as well as the finite quarkonium mass correction in $M_p/M_V$. In addition, the systematical uncertainties from the large $\xi$ expansion also require proper treatment.

Acknowledgment: We thank F. Yuan and L. Elouadrhiri for useful discussions and correspondences. We particularly thank Z.-E. Meziani for sharing the  data of the recent $J/\psi$-007 experiment. This research is partly supported by the U.S. Department of Energy, Office of Nuclear Physics, under contract number DE-SC0020682.

\bibliographystyle{apsrev4-2}
\bibliography{refs.bib}
\end{document}